# STAR FORMATION RATE IN LATE-TYPE GALAXIES: I- THE $H_\alpha$ AND FUV INTEGRATED VALUES.

M. A. Magaña Serrano,[1] A. M. Hidalgo Gámez,[1] I. Vega-Acevedo,[1] H.O. Castañeda[1]

*Draft version: November 12, 2019*

RESUMEN

Se obtuvo la Tasa de Formación Estelar (SFR) de una muestra de 36 galaxias tardías (14 dS y 22 Sm), a partir del flujo $H_\alpha$ y FUV. Se ha encontrado que la SFR(FUV) en la mayoría de los casos es mayor que la SFR($H_\alpha$) y que las galaxias barradas siempre presentan valores más pequeños de formación estelar que las que no tienen barra. También se obtuvo que las galaxias Sm tiene mayor formación estelar que las dS. Además, estudiamos la distribución espacial de la formación estelar dentro de nuestra muestra de galaxias, concluyendo que la mayoría de las galaxias son simétricas en su formación estelar, teniendo unos valores similares del número de regiones H II y del flujo de las mismas a ambos lados del eje de simetría. Finalmente, analizamos el papel que juega el Gas Ionizado Difuso en el valor de la SFR, encontrando que la luminosidad $H_\alpha$ proveniente de este gas no debería ser incluida en la determinación de la SFR a menos que se verifique que la pérdida de fotones ionizantes son los responsables de la ionización.

ABSTRACT

The Star Formation Rate (SFR) was determined from a sample of 36 late-type galaxies (14 dS and 22 Sm), from the $H_\alpha$ and Far Ultraviolet (FUV) flux. We found that the SFR(FUV) was in most cases higher than the SFR($H_\alpha$). We also obtained that the SFR is larger for Sm galaxies and smaller for barred galaxies, for any morphological type, in both diagnostic methods. In addition, a study of the spatial distribution of star formation within these galaxies was made, concluding that there is not a preferential place for the star formation. Finally, we studied the role of the Diffuse Ionized Gas in the SFR value, finding that the $H_\alpha$ flux contribution coming from this gas should not be included in the SFR determination unless it is verified that leaking photons are the only source of the neutral gas ionization

## 1. INTRODUCTION

Although late-type galaxies are relatively simple systems and are the most abundant galaxies in the Universe (Marzke & Da Costa, 1997; Mateo, 1998;

---

[1]Departamento de Física, Escuela Superior de Física y Matemáticas, Instituto Politécnico Nacional, U.P. Adolfo López Mateos,C.P. 07738, Mexico City, Mexico

Danieli, S. et al. 2017), their formation and evolution are not yet well understood. Although there are many different parameters which might influence the evolution of galaxies, one of the most important is the Star Formation Rate. The reasons for such importance are many. The SFR indicates how many stars are formed by the gas in a galaxy (in time). Therefore, gas mass is transformed into stars, which processes this gas transforming the primordial hydrogen into more complex atoms, like oxygen, sulfur or ion (see, for example, the seminal work of Tinsley on chemical evolution; Tinsley 1981). Moreover, massive stars formed are going to end their lives as supernovae, which inject about 4 $10^{42}$ erg of kinetic energy into the interstellar medium (ISM) of the galaxy (García-Segura et al. 1996). Such energy can be used to collapse the gas in the vicinity and form new stars (as in Holmberg II or IC 2574, Sánchez-Salcedo, 2002), to heat the medium avoiding a new event of star formation (Jog, 2013) or even to eject gas (and metals) outside the galaxy (e.g. Dekel & Silk, 1986; D'Ercole & Brighenti, 1999; Melioli et al. 2015). In order to have a wider and more comprehensive view of such events, other parameters are needed, such as the Initial Mass Function (IMF) which gives the number of stars formed for each mass interval, and the Star Formation History (SFH), which is the number of stars formed along the cosmic time for each galaxy. This set of parameters (SFR, IMF and SFH) provide information about the evolution of galaxies (López et al. 2018).

The SFR can be determined from several diagnostic methods. Two of the most important are the intensity of the recombination line $H_\alpha$ and the flux of the FUV (Kennicutt, 1998; Calzetti 2013; Audcent-Ross et al 2018). The $H_\alpha$ line has been widely used due to its many advantages, such as being easy to measure even with small aperture telescopes because of its large flux in star-forming galaxies. Nevertheless, for galaxies with $z > 0.5$, the $H_\alpha$ line goes into the Near Infrared and it is necessary to use a detector in this wavelength range. The UV measurements also allow the determination of the SFR from the FUV continuum. Several authors have noticed the differences in the SFR obtained from the $H_\alpha$ line and the FUV continuum and have given possible explanations for the discrepancies (e.g. Bell & Kennicutt, 2001; Lee et al. 2009): an IMF systematically deficient in the highest mass stars, a leakage of ionizing photons from a low density environment, or the fact that most methods only consider a single, solar metallicity. Most of these studies have been done on spiral and irregular galaxies of the local volume and for dwarf galaxies within 11 Mpc (Karachentsev & Kaisina, 2013; Lee et al. 2009).

It is well known that the highest star formation rates occur in Sc galaxies (Kennicutt, 1998). Late-type galaxies (Sm and dS) have a very large reservoir of neutral gas relative to their total mass (typically of the order of 6%, Huchtmeier & Richter, 1989) but a lower SFR (e.g. Hunter & Elmegreen, 2004). Therefore, the efficiency of the star formation process is very show for these galaxies. There could be several reasons for this, being the lack of a clear triggering mechanism of the star formation one of the most important ones.

There are several goals in this paper, which is focused on the integrated

SFR values of a sample of late-type galaxies. Firstly, to increase the sample of late-type galaxies with SFR measurements with $H_\alpha$ and FUV flux. Although there are many investigations devoted to SFR determinations (among others, James et al. 2004; Lee et al. 2009; Hunter & Elmegreen 2004; Hunter et al., 2010; Rosenberg et al. 2008; Buat et al. 2009; Boselli et al. 2015; Boselli et al. 2009; James et al. 2008; Almoznino & Brosch, 1996) there are only 50 Sm galaxies with SFR determined from the $FUV$ continuum, so the inclusion of another 28 galaxies increases the total sample to more than 50%. The raise is not so large for Sm galaxies with SFR from the recombination line $H_\alpha$: only about a 15%.

Another goal is to check how much the dS galaxies differ from Sm galaxies. Among other things, we can check the efficiency of the SFR, because the dS galaxies have a larger amount of neutral gas but a less clear spiral structure, as stated above.

Another interesting study performed is the spatial distribution of the H II regions in the galaxies. Some authors claim that most of the regions are located in the central part of the galaxies (e.g Roye & Hunter, 2000; Hodge, 1969; Bruch et al. 1998), and that, for barred galaxies, many H II regions are located at the end of the bar (Elmegreen & Elmegreen, 1980). This is tested for the galaxies in our sample. Moreover, if there is any large asymmetry in the H II regions distribution, some clues on the recent environmental history can be obtained.

Finally, we want to study how the diffuse ionized gas (hereafter, DIG) might affect the estimation of the SFR in galaxies, because not all $H_\alpha$ emission outside the H II regions is due to ionizing photons emitted by massive stars, but also by other mechanisms (See Hidalgo-Gámez 2004 for a review). So, the inclusion of DIG photons in the total $H\alpha$ emission might give a higher, but not so accurate SFR.

The paper is organized as follows. In Section 2 we describe the galaxy sample, the data acquisition, reduction and calibration of data as well as detection of the H II regions. The determination of the SFR values with the $H_\alpha$ and FUV fluxes, along with a contrast of our results with the data from the literature and the efficiency of the process of the star formations are studied in Section 3, while the results on the distribution of the H II regions in our sample is outlined in Section 4. Finally, the importance of the DIG on the estimation of the SFR is presented in Section 5. Conclusions are presented in Section 6.

## 2. SAMPLE DESCRIPTION, DATA ACQUISITION AND REDUCTION

The sample used in this investigation consists of 36 late-type galaxies, that are divided into 14 dwarf spiral galaxies (dS) and 22 Sm galaxies. The galaxies in the latter subsample were selected from the RC3 and UGC catalogues, considering only Sm or types 9 or 10 galaxies. The galaxies in the former subsample were selected from Table 1 in Hidalgo-Gámez (2004). All the interacting and active galaxies were disregarded. Ten of the galaxies in this sample are classified as barred, 7 of them are Sm and 3 dS. Finally, from a visual inspection of the H$_\alpha$ images we classified our sample into galaxies with clear spiral structure (18 galaxies), galaxies without spiral structure (9 galaxies) and those of intermediate type (9 galaxies). The galaxies in the sample presented here were selected from the tables mentioned above and observed just because their coordinates were the right ones at the moment of the observations.

The number of Sm galaxies studied here is comparable with other samples [8 Sm galaxies in James et al. (2004), 20 Sm in Hunter & Elmegreen (2004), 25 Sm in van Zee (2001) and 7 Sm galaxies in Hunter et al. (2010)].

The characteristics of the galaxies in the sample are presented in Table 1. The name and the morphological type is presented in columns 1 and 2, while the coordinates ($\alpha$ and $\delta$) are listed in columns 3 and 4 from NED-NASA. The absolute magnitude (column 5) was determined from the apparent magnitude and the kinematic distance (column 7) which were obtained from NASA Extragalactic Database (NED-NASA). The absolute magnitudes are galactic extinction corrected (Column 8). The inclination is presented in column 9 and was also obtained from NED-NASA. The optical size (Column 6) and the surface brightness (column 10) were determined following equations 3 and 4 in Hidalgo-Gámez & Olofsson (1998). Finally, the H I mass and surface density were computed from equation 2 in Hidalgo-Gámez (2004).

### 2.1. *Observations, reduction and calibration of the data*

The H$_\alpha$ images were acquired at the 1.5m telescope of the Observatorio Astronómico Nacional at San Pedro Mártir (OAN-SPM) in five different observational campaigns from 2002 to 2004. Five narrow band filters were used: three in the H$_{H\alpha}$ recombination line, centered at 6570 Å, 6607 Å and 6690 Å, and two for continuum subtraction, with central wavelengths at 6459 Å and 6450 Å. The maximum transmittance of the filter was 70% and they were between 89 Å and 127 Å wide. The integration times were in the range of 1200 and 5400$s$ for H$_{H\alpha}$ and 900 and 3600$s$ for continuum, respectively. The air mass was lower than 1.4 for all the galaxies but three, and the seeing was between 1.1 and 2.6, with an average value of 1.4. The convolution of the peak transmission of the filters (70%) and the high detector quantum efficiency (90%) allowed to achieve relatively deep flux limits ($7 \times 10^{-18} erg cm^{-2} s^{-1}$ ).

The reduction procedure (bias, flat-fields, removal of cosmic rays and sky subtraction) as well as the flux calibration were performed using the ESO-MIDAS software. Three standard stars were observed each night with different air mass in order to perform the flux calibration.

The ultraviolet fluxes were obtained from the 6th release (GR6) of the GALEX database for a total of 30 out of the 36 galaxies of the sample. We used those data with exposure time of 1500s. Both the FUV and NUV fluxes were determined for the 30 galaxies from the integrated flux provided by the conversion from AB magnitudes: $F_v[erg\ s^{-1}\ cm^{-2}\ Hz^{-1}] = 10^{-0.4(m_{AB}+48.6)}$.

TABLE 1

GLOBAL PARAMETERS

| Galaxy | Type | $\alpha$ 2000 | $\delta$ 2000 | $M_B$ [mag] | $r_{25}$ [kpc] | D [Mpc] | $A_R$ [mag] | i [°] | $\mu_B$ [$mag/arcsec^2$] | M(HI) [$M_\odot * 10^9$] | $\Sigma_{HI}$ [$M_\odot pc^2$] | $log(M_{H_I}/L_B)$ |
|---|---|---|---|---|---|---|---|---|---|---|---|---|
| DDO 18 | Sm | 02 10 44 | 06 45 30 | -18.37 | 6.59 | 20.70 | 0.13 | 70 | 23.8 | 0.99 | 5.57 | -1.72 |
| UGC 560 | Sm | 00 54 47 | 13 39 29 | -20.5 | 10.57 | 72.70 | 0.21 | 24 | 26.1 | – | – | – |
| UGC 3086 | Sm | 04 32 55 | 00 32 12 | -16.64 | 9.75 | 73.50 | 0.18 | – | – | 4.61 | 12.85 | -1.45 |
| UGC 3778 | Sm | 07 16 55 | 28 31 46 | -17.97 | 9.57 | 65.80 | 0.14 | – | – | 4.1 | 11.44 | -1.94 |
| UGC 3947 | Sm | 07 39 02 | 33 54 59 | -17.12 | 7.41 | 55.90 | 0.10 | 45 | 24.7 | 1.35 | 9.20 | -1.94 |
| UGC 3989 | Sm | 07 44 41 | 53 50 05 | -17.76 | 11.56 | 79.50 | 0.11 | 0 | 26.3 | – | 9.21 | – |
| UGC 4121 | Sm | 07 58 54 | 54 02 33 | -17.55 | 6.44 | 18.90 | 0.08 | 72 | 24.3 | 1.2 | 6.69 | -1.23 |
| UGC 4797 | Sm | 09 08 11 | 05 55 40 | -17.41 | 5.43 | 19.60 | 0.10 | 24 | 24.2 | 0.62 | – | -1.5 |
| UGC 4837 | Sm | 09 21 10 | 35 31 54 | -17.50 | 8.78 | 28.90 | 0.06 | 51 | 24.5 | 1.84 | 14.46 | -1.4 |
| UGC 5236 | Sm | 09 47 00 | 21 43 47 | -20.54 | 10.00 | 53.40 | 0.08 | 12 | 25.3 | 0.03 | 0.09 | -4.92 |
| UGC 6151 | Sm | 11 05 56 | 11 49 35 | -17.29 | 6.73 | 24.30 | 0.06 | 45 | 24.3 | 0.88 | 4.74 | -1.47 |
| UGC 6205 | Sm | 11 09 59 | 46 05 44 | -19.58 | 4.99 | 27.30 | 0.04 | 41 | 23.7 | 0.31 | 3.96 | -2.94 |
| UGC 6399 | Sm | 11 23 23 | 50 53 34 | -18.02 | 8.13 | 20.30 | 0.04 | 74 | 23.1 | 0.69 | 7.36 | -1.71 |
| UGC 8253 | Sm | 13 10 44 | 11 42 28 | -17.72 | 11.16 | 53.10 | 0.07 | 17 | 25.4 | 2.83 | – | -1.82 |
| DDO 36 | SBm | 05 07 47 | -16 17 37 | -18.74 | 8.49 | 24.90 | 0.17 | 54 | 23.1 | – | – | – |
| NGC 2552 | Sm | 08 19 20 | 50 00 35 | -19.37 | 5.14 | 10.20 | 0.12 | 49 | 23.5 | 0.74 | 8.62 | -1.63 |
| NGC 4010 | SBm | 11 58 38 | 47 15 41 | -20.47 | 11.29 | 18.20 | 0.05 | 79 | 22.9 | 2.57 | 16.1 | -2.0 |
| UGC 4871 | SBm | 09 14 57 | 39 15 45 | -17.79 | 9.14 | 37.00 | 0.03 | 69 | 23.7 | – | 6.18 | – |
| UGC 8385 | SBm | 13 20 38 | 09 47 14 | -19.70 | 7.29 | 22.40 | 0.05 | 56 | 23.7 | – | – | – |
| UGC 10058 | SBm | 15 50 24 | 25 55 21 | -17.99 | 5.49 | 34.4 | 0.14 | 42 | 25.3 | – | – | – |
| UGCA 117 | SBm | 06 00 35 | -28 59 31 | -19.53 | 11.50 | 30.75 | 0.08 | 56 | 23.0 | 1.4 | 11.86 | -2.37 |
| UGC 2345 | SBm | 02 51 53 | -01 10 20 | -16.23 | 7.16 | 14.20 | 0.19 | 29 | 27.0 | – | – | – |
| UGC 3775 | dS | 07 15 53 | 12 06 54 | -17.34 | 4.68 | 29.40 | 0.24 | 27 | 25.3 | 0.41 | 5.96 | -1.99 |
| UGC 4660 | dS | 08 54 24 | 34 33 22 | -15.96 | 5.57 | 32.60 | 0.06 | 21 | 25.5 | 1.5 | 8.52 | -0.96 |
| UGC 5296 | dS | 09 53 11 | 58 28 42 | -16.06 | 3.02 | 25.20 | 0.03 | 29 | 24.5 | 0.29 | 10.12 | -1.5 |
| UGC 5740 | dS | 10 34 46 | 50 46 06 | -16.63 | 2.86 | 11.30 | 0.05 | 46 | 24.6 | 0.41 | 6.31 | -0.87 |
| UGC 6304 | dS | 11 17 49 | 58 21 05 | -16.40 | 4.72 | 28.90 | 0.03 | 39 | 25.0 | – | – | – |
| UGC 6713 | dS | 11 44 25 | 48 50 07 | -16.68 | 3.66 | 17.00 | 0.05 | 44 | 24.0 | 0.9 | 13.78 | -0.9 |
| UGC 9018 | dS | 14 05 33 | 54 27 40 | -14.95 | 1.59 | 6.58 | 0.03 | 37 | 23.7 | 0.13 | 13.78 | -0.23 |
| UGC 9902 | dS | 15 34 33 | 15 08 00 | -15.93 | 3.67 | 27.70 | 0.13 | 72 | 23.5 | – | – | – |
| UGC 9570 | dS | 14 51 36 | 58 57 14 | -17.46 | – | 35.50 | 0.03 | – | – | – | – | – |
| UGC 11820 | dS | 21 49 28 | 14 13 52 | -15.79 | 4.96 | 17.10 | 0.33 | 24 | 25.5 | 1.54 | 19.96 | -0.32 |
| UGC 12212 | dS | 22 50 30 | 29 08 18 | -20.77 | 3.30 | 19.50 | 0.17 | 56 | 24.1 | 0.8 | 23.38 | -2.72 |
| UGC 891 | dSB | 01 21 19 | 12 24 43 | -17.30 | 3.13 | 9.40 | 0.08 | 63 | 24.1 | 0.36 | 11.70 | -1.04 |
| UGC 5242 | dSB | 09 47 06 | 00 57 51 | -20.57 | 4.86 | 26.30 | 0.34 | 50 | 24.0 | 0.8 | 10.78 | -2.9 |
| UGC 6840 | dSB | 11 52 07 | 52 06 29 | -16.99 | 4.60 | 17.00 | 0.06 | 72 | 22.9 | 1.04 | 12.77 | -0.97 |

Some characteristics of the galaxies in the sample are listed. The morphological type is shown in column 2. The coordinates (2000) are given in columns 3 and 4. The absolute magnitude and the optical size are given in columns 5 and 6. The kinematic distance ($V_{LG}/H_o$) is included in column 7. $A_R$ extinction is listed in column 8. The inclination is listed in column 9. Surface brightness is given in column 10. The mass of hydrogen gas $M(HI)$ and the mass surface density are given in columns 11 and 12 respectively, and the mass luminosity ratio is shown in column 13.

Full details on the telescope, instruments, calibration and processing pipeline are provided by Martin et al. (2005) and Morrissey et al. (2005, 2007).

### 2.2. H II Regions detection

In order to obtain the SFR, fluxes from the H$_\alpha$ line are needed as well as in the FUV continuum. The simplest approach is to consider all galaxy emission as a whole, e.g. the continuum-subtracted H$_\alpha$ emission. However, as we were interested in a further analysis of the data (such as the Luminosity Function and the location of the H II regions inside the galaxies) we decided not to use that technique in our work. Instead, we selected the H II regions from the continuum-subtracted H$_\alpha$ images. Only those regions identified by two of us were included in the sample. A circular aperture centered on each region was used to determine the flux. Two different radii were considered for each H II region. The first one included all the emission of the region and the second one included only the emission with flux values larger than a limiting value, in order to distinguish between proper H II regions and DIG. As will be discussed in Section 6, such limiting flux has been determined from spectroscopic studies and is of $10^{-17}$ erg $s^{-1}$ cm$^{-2}$. This procedure was performed manually, and the radius aperture just depends on galaxy distance, which define the size of the region in the image.

The main disadvantage of this procedure is the difficulty to detect very low luminosity regions. However, as they are very faint, the contribution of their fluxes to the SFR is not significant and will only be important in the study of the luminosity function of the H II regions inside the galaxies.

In the case of galaxies with large inclinations, it is very difficult to distinguish different regions if they are in the same line-of-sight. This is the situation for 7 of the galaxies in our sample, which have inclinations larger than 70$^o$. The inclusion of such galaxies in our sample can be argued because of the difficulty of an accurate extinction correction. However, most of the classical studies of the SFR in galaxies included high inclination ones without any further correction (e.g. James et a. 2008; Lee et al. 2099; Hunter et al. 2010; Boselli et al. 2015). In 2018, Wang et al. studied the influence of the inclination in the SFR value and they concluded that the inclination is important only for galaxies more massive than $10^{10} M_\odot$. Moreover, Buat et al. (2009) considered high inclination galaxies only those with inclination values larger than 80$^o$. As the galaxies in our sample are less massive than $10^{10} M_\odot$ and have inclinations smaller than 80$^o$, we think that the values determined here are accurate enough although they should be considered as upper limits in the SFR.

A similar problem might appear for those H II regions which are very close and cannot be separated due to the resolution of our observations.

With the knowledge of the positions of the H II regions from the H$_\alpha$ images, the GALEX images were inspected to locate the star forming region and the fluxes from these locations were calculated.

### 2.3. *Flux corrections*

Before the SFR could be determined, the UV and H$_\alpha$ fluxes need to be corrected by extinction. There are contributions to the extinction from the gas inside the galaxy (hereafter, the internal extinction): ii) due to the Intergalactic Medium, iii) due to gas inside our own Galaxy, and iv) due to the atmosphere. The last one was performed in the standard way. To correct for Galactic extinction we followed Cardelli et al. (1989), using values of $R_V$ from 2.6 to 5.5 (Clayton & Cardelli, 1988) and considering the colour excess values from NED.

The most difficult extinction correction to estimate is the extinction internal to the galaxy. Due to the lack of a simple way to estimate the extinction coefficient in $H_\alpha$, $A(H_\alpha)$, this correction is not generally computed. In this article, the determination of the internal extinction coefficient for $H_\alpha$ was computed following Calzetti et al. (2000), where they derived an extinction law $K(\lambda)$ directly from the data in UV and optical wavelength range, as in Calzetti (1997b). As $K(\lambda)$ is related to the internal nebular extinction, the extinction coefficient in $H_\alpha$ can be obtained as:

$$A(H_\alpha) = K(\lambda) \times E(B-V)$$

On the other hand, we have followed Salim et al (2007)'s formalism to estimate the extinction coefficient in FUV, $A_{FUV}$, as

$$A_{FUV} = \begin{cases} 3.32(m^0_{FUV} - m^0_{NUV}) + 0.22, & \text{if } (m^0_{FUV} - m^0_{NUV}) < 0.95. \\ 3.32, & \text{if } (m^0_{FUV} - m^0_{NUV}) > 0.95. \end{cases}$$

where $m^0_{FUV} - m^0_{NUV}$ is the UV colour of the galaxy in the rest-frame system. The corrected fluxes obtained were of the order of $10^{-26}$ $erg$ $cm^{-2}$ $s^{-1}$ with typical values of $A_{FUV}$ between 0.1 to 2.5 mag.

The internal extinction increased the final fluxes in the galaxies of our sample on an average 15%, but the increment was up to 50% for three of the galaxies (UGC 560, UGC 5242, UGC 12212).

Because of the large bandwidth of the filters used in this work (larger than 70 Å), the $H_\alpha$ flux is contaminated with the nitrogen lines emission. From our own spectroscopy data of several of the galaxies presented here, the $[NII]/H_\alpha$ ratio is typically smaller than 0.1 (Hidalgo-Gámez et al. 2012). The flux obtained from our images was then corrected by 10% in order to eliminate this contribution.

## 3. INTEGRATED STAR FORMATION RATES

The star formation rates were determined from the corrected $H_\alpha$ and FUV fluxes using the Kennicutt's (1998) expressions. The total flux is the sum of the fluxes for all the H II regions. Recently, Hunter et al. (2010) obtained a new SFR calibration for low metallicity galaxies. There are differences of about 15% in the SFR values determined with Hunter's or Kennicutt's expressions. Although the galaxies in our sample have subsolar metallicity (Hidalgo-Gámez et al., in preparation; Hidalgo-Gámez et al. 2012) we used Kennicutt's because the comparison with previous work is straightforward. The SFR values determined with Hunter et al.'s expressions are listed in columns 7 ($H_\alpha$) and 9 (UV) of Table 2 for a quick comparison with Kennicutt's ones. The expressions used by Kennicutt (1998) are

$$SFR[M_\odot \ yr^{-1}] = 9.93 \ 10^{-41} F(H_\alpha) \cdot D^2$$

$$SFR[M_\odot \ yr^{-1}] = 1.4 \ 10^{-28} L_{FUV}(erg s^{-1} Hz^{-1})$$

The distance used to determine the luminosities is the so-called kinematic distance, $D = V_{LG}/H_o$, corrected by Virgo Cluster infall, where $H_o$ is 73 km s$^{-1}$ Mpc$^{-1}$ (Riess et al. 2011).

The SFRs in H$_\alpha$ and FUV for all galaxies, separated by morphological type, as well as the total fluxes and the extinction values for both bands, are shown in Table 2. Two different fluxes were considered for H$_\alpha$, with and without DIG. The latter values are shown in the last column of Table 2 (see Section 6 for details). The errors were determined considering the uncertainties in the distance and the error associated with the flux determination.

### 3.1. Star Formation Rates of late-type galaxies from H$_\alpha$ luminosities

From a closer inspection of column 6 in Table 2, it can be concluded that there are three ranges in the SFR($H_\alpha$) values: galaxies with values larger than 1 M$_\odot$ $yr^{-1}$ (3 galaxies), those with SFR($H_\alpha$) between 0.1 and 1 M$_\odot$ $yr^{-1}$ (9 galaxies), and galaxies with values smaller than 0.1 M$_\odot$ $yr^{-1}$ (24 galaxies). For the galaxy UGC 9902 the SFR is taken as zero because we could not detect any H$_\alpha$ flux in it. The distribution of the SFR($H_\alpha$) values for all the galaxies in the sample is shown in Figure 1 (left). The distribution is not symmetrical with a clear asymmetry towards the low SFR values.

The galaxy with the largest SFR($H_\alpha$) value is UGC 560, one of the most distant ones in our sample, with a SFR similar to the Milky Way (MW) rate ($1.7 M_\odot \ yr^{-1}$, Robitaille & Whitney, 2010) while the one with the smallest SFR is UGC 9018, a dS galaxy which is the closest in the sample, one of the smallest in size and with very few gas left.

As can be seen from figure 2 (left column), where the distribution for Sm galaxies is shown at the top row while the one for dS galaxies is at the bottom, the former have larger SFR values than the dS galaxies, a factor of 4.7 according to the average values shown in Table 3 (second column). Actually, only one out of 14 dS galaxies has a SFR larger than 0.1 M$_\odot$ $yr^{-1}$. One explanation might be the small number of dS galaxies in our sample. However, they are almost a 40% of the total sample and, therefore, such differences might be due to another reason. We will further discuss this in Section 3.6

### 3.2. Star Formation Rates of late-type galaxies from UV continuum

The most interesting feature of the SFR($FUV$) values, listed in column eighth of Table 2 and shown in Figure 1 (right), is the lack of values larger than 1M$_\odot$ $yr^{-1}$. The galaxy with the largest value is UGC 5236, an almost face-on Sm galaxy, with SFR(FUV)= 0.76 $M_\odot \ yr^{-1}$. Interestingly, the galaxy with the lowest SFR, UGC 8253, is very similar to UGC 5236 in size, inclination and distance. The SFR distribution is not symmetrical as well, with a deficit of galaxies with low SFRs.

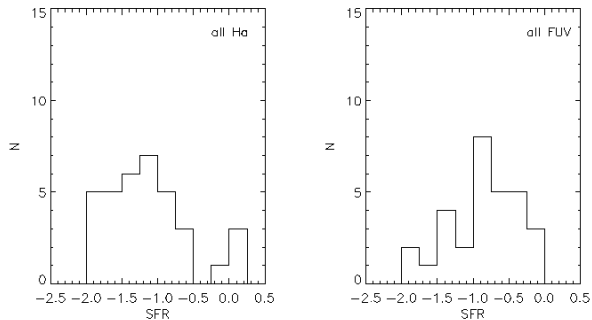

Fig. 1. Star formation rate distributions for all the galaxies in the sample from H$\alpha$ fluxes with DIG (left) and from FUV continuum (right). The range comprised for the H$\alpha$ fluxes is wider than for the FUV, with the bulk of galaxies below 0.1 M$_\odot$ $yr^{-1}$ for the former and between 1 and 0.1 for the latter.

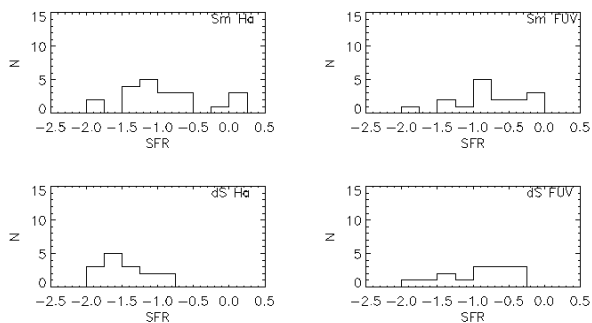

Fig. 2. Star formation rate distribution for the Sm galaxies (upper row) and dS galaxies (bottom row). The left column shows the values with the $H\alpha$ line ($SFR(H_\alpha)_{DIG}$), while the values with the FUV are plotted in the right column.

Twenty one out of 30 galaxies have SFR(FUV) values between 1 and $0.1 M_\odot$ $yr^{-1}$, (14 Sm and 9 dS) and only 9 galaxies have values lower than $0.1 M_\odot$ $yr^{-1}$. There are differences of 0.11 between the SFR($FUV$) of Sm and dS. Although the distribution has a similar range, as can be seen in figure 2, Sm galaxies have a peak at 0.1 $M_\odot$ $yr^{-1}$ while dS galaxies have a smooth distribution.

### 3.3. Comparison between SFR(H$_\alpha$) and SFR(UV)

It is well known that SFR($FUV$) is larger than the values obtained using the $H_\alpha$ luminosity (Bell & Kennicutt, 2001; Lee et al. 2004; Lee et al. 2009). The UV emission comes from O and B stars with masses larger

TABLE 2

$H_\alpha$ AND $FUV$ STAR FORMATION RATE

| Galaxy | $F_{H\alpha}$ $10^{-14}$ [erg s$^{-1}$ cm$^{-2}$] | $A_{H\alpha}$ [mag] | $F_{FUV}$ $10^{-26}$ [erg s$^{-1}$ Hz$^{-1}$] | $A_{FUV}$ [mag] | SFR($H_\alpha$)$_{DIG}$ [$M_\odot yr^{-1}$] $10^{-2}$ | SFR($H_\alpha$)$_{Hunter}$ [$M_\odot yr^{-1}$] $10^{-2}$ | SFR($FUV$) [$M_\odot yr^{-1}$] $10^{-2}$ | SFR($FUV$)$_{Hunter}$ [$M_\odot yr^{-1}$] $10^{-2}$ | SFR($H_\alpha$) [$M_\odot yr^{-1}$] $10^{-2}$ |
|---|---|---|---|---|---|---|---|---|---|
| DDO 18 | 9.12 | 0.28 | 0.95 | 0.78 | 17±2 | 14.8 | 14± 2 | 12.70 | 3.7± 0.5 |
| UGC 560 | 11.13 | 0.77 | 0.10 | 2.11 | 113±17 | 97.78 | 63± 10 | 57.14 | 56 ± 8 |
| UGC 3086 | 19.10 | – | – | – | 109±5 | 95.2 | – | – | 97 ± 3 |
| UGC 3778 | 6.04 | – | – | – | 108 ± 16 | 94.28 | – | – | 25 ± 3 |
| UGC 3947 | 1.01 | – | – | – | 5.8±0.9 | 5.1 | – | – | 2.9± 0.6 |
| UGC 3989 | 1.23 | 0.19 | 0.43 | 0.52 | 66 ±2 | 57.6 | 74 ± 13 | 67.12 | 7.3± 0.4 |
| UGC 4121 | 1.19 | 0.25 | 0.55 | 0.69 | 3.7 ± 0.8 | 3.2 | 6.0± 1.0 | 5.44 | 0.4± 0.1 |
| UGC 4797 | 1.13 | 0.05 | 2.33 | 0.13 | 4.1±0.7 | 3.6 | 17 ± 2 | 15.42 | 0.41± 0.07 |
| UGC 4837 | 7.02 | 0.00 | 0.72 | 0.01 | 19±3 | 17 | 10± 4 | 9.07 | 5.5 ± 0.7 |
| UGC 5236 | 6.63 | 0.62 | 0.33 | 1.70 | 18±8 | 16 | 76 ± 14 | 68.93 | 18 ± 4 |
| UGC 6151 | 1.20 | 0.01 | 1.21 | 0.03 | 7.1±0.9 | 6.2 | 12± 2 | 10.88 | 0.67± 0.08 |
| UGC 6205 | 5.23 | 0.31 | 0.90 | 0.85 | 4 ± 1 | 3 | 25± 4 | 22.68 | 4 ±1 |
| UGC 6399 | 0.45 | 0.04 | 0.56 | 0.10 | 1.5±0.4 | 1.3 | 4 ± 1 | 3.63 | 0.18± 0.04 |
| UGC 8253 | 1.56 | 0.03 | 0.03 | 0.09 | 4.3±0.2 | 3.8 | 1.5± 0.5 | 1.36 | 4.2± 0.2 |
| DDO 36 | 21.94 | – | – | – | 69 ±10 | 60 | – | – | 13± 2 |
| NGC 2552 | 103.89 | 0.34 | 9.72 | 0.92 | 15 ±4 | 13 | 39± 4 | 35.37 | 10 ± 3 |
| NGC 4010 | 1.50 | 0.42 | 1.30 | 1.16 | 5.8±0.6 | 5.1 | 21 ± 6 | 19.05 | 0.47± 0.07 |
| UGC 4871 | 1.39 | 0.00 | 0.61 | 0.01 | 15.4±0.4 | 13.4 | 14 ± 2 | 12.69 | 1.80± 0.05 |
| UGC 8385 | 8.03 | 0.39 | 2.03 | 1.07 | 6.0±0.3 | 5.2 | 45.7 | 41.45 | 3.8± 0.3 |
| UGC 10058 | 0.07 | 0.60 | 0.22 | 0.08 | 1.5± 0.3 | 1.3 | 3.7± 0.5 | 3.36 | 2.5±0.5 |
| UGCA 117 | 0.54 | 0.01 | 1.91 | 0.04 | 4.9±0.7 | 4.3 | 31± 5 | 28.12 | 0.49± 0.06 |
| UGC 2345 | 29.73 | 0.27 | 5.03 | 0.73 | 12±1 | 10 | 33± 4 | 29.9 | 5.7± 0.6 |
| UGC 3775 | 2.88 | 0.16 | 0.69 | 0.45 | 6 ± 1 | 5 | 15 ± 2 | 13.6 | 2.3±0.9 |
| UGC 4660 | 4.24 | – | – | – | 6±1 | 5 | – | – | 4.3± 0.6 |
| UGC 5296 | 2.56 | – | – | – | 2.3± 0.2 | 2.0 | – | – | 1.5±0.1 |
| UGC 5740 | 4.05 | 0.31 | 1.57 | 0.86 | 2.1±0.3 | 1.8 | 7 ± 1 | 6.35 | 0.49± 0.06 |
| UGC 6304 | 3.74 | 0.01 | 2.06 | 0.04 | 5.6± 0.1 | 4.9 | 30± 4 | 27.21 | 2.95±0.07 |
| UGC 6713 | 4.47 | 0.09 | 1.65 | 0.25 | 1.3±0.3 | 1.1 | 10± 4 | 9.07 | 1.2± 0.3 |
| UGC 9018 | 7.72 | 0.02 | 5.52 | 0.05 | 1.0±0.1 | 0.9 | 4.2±0.6 | 3.81 | 0.32± 0.04 |
| UGC 9902 | 0.00 | 0.01 | 0.13 | 0.02 | 0.00 | 0.00 | 1.7± 0.3 | 1.54 | 0.00 |
| UGC 9570 | 0.19 | 0.15 | 1.52 | 0.40 | 2.3± 0.1 | 2.0 | 46± 6 | 41.72 | 0.23±0.01 |
| UGC 11820 | 5.30 | 0.02 | 0.45 | 0.05 | 1.7± 0.4 | 1.5 | 2.3 | 2.09 | 1.5± 0.3 |
| UGC 12212 | 20.23 | 0.91 | 0.67 | 2.50 | 14±2 | 12 | 42±6 | 38.1 | 7.3±0.9 |
| UGC 891 | 8.27 | 0.41 | 2.51 | 1.12 | 2.6± 0.4 | 2.3 | 10 ± 1 | 9.07 | 0.70±0.3 |
| UGC 5242 | 2.97 | 0.76 | 0.24 | 2.07 | 6.0± 0.8 | 5.2 | 19±3 | 17.23 | 1.9±0.2 |
| UGC 6840 | 2.79 | 0.01 | 0.74 | 0.02 | 2.01±0.02 | 1.75 | 4± 2 | 3.63 | 0.76± 0.02 |

The $H_\alpha$ and $FUV$ fluxes are given in columns 2 and 4. The internal dust extinction $A_{H\alpha}$ and $A_{FUV}$ are listed in columns 3 and 5. The SFR($H_\alpha$) with DIG, and the SFR($H_\alpha$) that is obtained from the Hunter et al.'s calibration are shown in columns 6 and 7. The SFR($FUV$) from Kennicutt and Hunter et al.'s calibrations are listed in columns 8 and 9, respectively, and the SFR($H_\alpha$) without DIG are shown in column 10.

than 5 M$_\odot$, while only stars more massive than 20 M$_\odot$ can provide the H$_\alpha$ emission (Werk et al. 2010). This can be checked in Figure 3, where the $SFR(FUV)/SFR(H_\alpha)$ ratio vs. the SFR($H_\alpha$) values are plotted. Only five galaxies, all of them Sm, are in the negative locus of the diagram. The lower the SFR($H_\alpha$), the higher the differences between both values; as in the earliest stages of the starburst, the emission in $H_\alpha$ is higher and both fluxes are similar. At latter stages of the star formation events, the differences between the fluxes in FUV and H$_\alpha$ become larger as the UV continuum decreases more slowly than H$_\alpha$. The main reason for this is that the UV photons are origi-

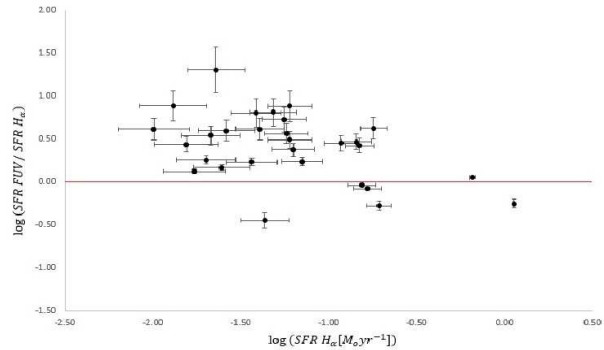

Fig. 3. Ratio of $SFR(FUV)/SFR(H_\alpha)$ vs. SFR($H_\alpha$). The solid line indicates when both values are the same. Most of the galaxies show a positive value for this ratio indicating that the SFR obtained with the UV continuum is larger than the SFR for $H_\alpha$ with DIG.

TABLE 3

AVERAGED STAR FORMATION RATE

|  | $H_\alpha$ | $\sigma$ | Without DIG | $\sigma$ | FUV | $\sigma$ |
|---|---|---|---|---|---|---|
| All | 0.19 | 0.2 | 0.08 | 0.3 | 0.23 | 0.3 |
| All Sm | 0.28 | 0.4 | 0.12 | 0.2 | 0.27 | 0.4 |
| All dS | 0.06 | 0.05 | 0.02 | 0.03 | 0.16 | 0.2 |
| Barred | 0.12 | 0.2 | 0.03 | 0.2 | 0.20 | 0.5 |
| non-barred | 0.22 | 0.4 | 0.10 | 0.4 | 0.24 | 0.2 |
| Barred Sm | 0.17 | 0.2 | 0.04 | 0.05 | 0.25 | 0.6 |
| non-barred Sm | 0.33 | 0.4 | 0.16 | 0.3 | 0.28 | 0.3 |
| Barred dS | 0.03 | 0.03 | 0.01 | 0.01 | 0.11 | 0.2 |
| non-barred dS | 0.04 | 0.05 | 0.02 | 0.4 | 0.18 | 0.2 |

Averaged Star Formation Rate with the $H\alpha$ line (left columns), the $H\alpha$ emission from the H II region only, (middle columns) and from the FUV continuum (right columns). The total sample was divided into several samples: Sm and dS (upper rows), barred and non-barred galaxies (middle rows), and barred/non barred Sm and dS (bottom rows). Two things can be noted from this table: the non-barred galaxies always have larger amount of star formation compared to the barred galaxies, and Sm galaxies have larger Star Formation Rates than dwarf galaxies.

nated from a wider range of stars in the main sequence with larger life-time (e.g. Sullivan et al. 2004; Iglesias-Páramo et al. 2004).

From the values of Table 3, the average star formation rates in $H\alpha$ and $FUV$ are very similar for the sample as a whole as well as for Sm galaxies, and there are only large differences for dS galaxies (a factor of 2.7). One explanation might be that the star formation event is older in the latter galaxies, therefore all the massive stars which can ionize hydrogen have disappeared already and the $H_\alpha$ flux is quite low. However, there are still a lot of less massive stars which can emit at FUV wavelengths, hence the large value of the SFR($FUV$).

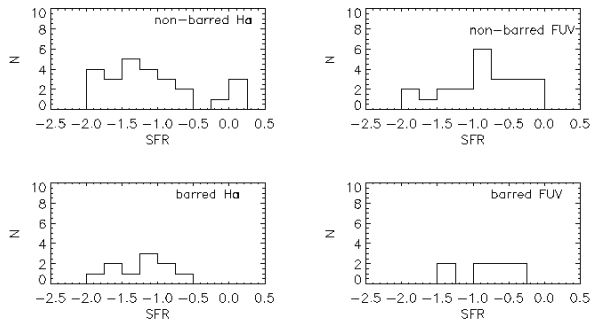

Fig. 4. Star formation rate distribution for barred (bottom row) and non-barred galaxies (top row) in the sample. The left column plots the SFR determined with the H$_\alpha$ luminosity while the right column shows the distribution of the SFR determined with the FUV continuum. Non-barred galaxies have, on average, larger Star Formation Rates.

### 3.4. *Bars and star formation rates*

It has been suggested that strong barred galaxies have larger SFRs (e.g. Sérsic & Pastoriza 1967; Ho et al. 1997; Tsai et al. 2013). However, other autors proposed that bars may decrease the SFR of a galaxy (e.g. Tubbs 1982; Kim et al. 2018). Kim et al. (2017), using a large sample of galaxies from SDSS, found out that barred galaxies have significantly lower star formation activity than their unbarred counterparts. Ryder & Dopita (1994) stated that the effect of the bar in the star formation rate is not important. Therefore, the problem is far from being settled yet.

We analized the effect of the existence of bars in the SFR of our sample. Barred galaxies represent a 28% of the galaxies in our sample, therefore the results are significant. Figure 4 shows the distribution of the SFR for barred and non-barred galaxies, with the average values presented in Table 3. Although non-barred galaxies have larger SFR than barred ones for $H_\alpha$ values, while the SFR is similar when determined with $FUV$ fluxes, the distributions are very different in both cases. Barred galaxies have a small range of SFR, with no large values neither with $H_\alpha$ nor with $FUV$. The distribution of barred galaxies with $FUV$ values is flat. From our sample it can be concluded that barred galaxies have lower SFR than non-barred galaxies.

How can this influence the lower values of SFR for dS? Actually, the number of barred dS galaxies is small, only three out of 14, therefore the results cannot be conclusive. However, there are no differences in SFR($H_\alpha$) while the differences increase for SFR($FUV$). So, the small SFR for dS galaxies are due to another reason.

### 3.5. *Comparison with previous results*

There are several studies of SFR involving late-type galaxies that use different diagnostic methods (e.g. Hunter & Elmegreen 2004; Hunter et al. 2010;

Rosenberg et al. 2008; Lee et al. 2009; Buat et al. 2009; Boselli et al. 2015; Boselli et al. 2009; James et al. 2008; Almoznino & Brosch, 1996). In general, Sm galaxies have higher values than Im galaxies in all these investigations, being the values for the Im closer to the SFR of our dS galaxies. On average, the SFR values reported for Sm galaxies are very similar to those we determined when extinction is considered. This is particularly important for the SFR($FUV$), because the extinction is larger than in the optical range. This might be the main reason for the differences between our SFR($FUV$) values and those reported by Lee et al. (2009) and Hunter & Elmegreen (2004), with the latter authors using colour excess and the Cardelli et al's extinction law, and the former ones having used Balmer decrement to determine the internal extinction. Therefore, our values are between two to five times larger than theirs.

Concerning the comparison between the SFR($H_\alpha$) and SFR($FUV$), our results are similar to other investigations, being the SFR($FUV$) higher than the SFR($H_\alpha$), except for the KISS sample by Rosenberg et al. (2008), where at least 8 out of 19 galaxies have lower SFR($FUV$).

Eleven of the galaxies studied here have SFR values previously determined from $H_\alpha$ emission (Hunter & Elmegree, 2004; James et al. 2004; Lee et al. 2009; Van Zee, 2001), and four have previous $FUV$ emission determinations (Hunter & Elmegree, 2010; Lee et al. 2009). Six of the values for SFR($H_\alpha$) are identical between our values and those previously determined, while another five are similar when the uncertainties are considered. The most discordant SFR($H_\alpha$) values are for UGC 9018, UGC 8385, DDO 18 and James et al.'s value for UGC 11820. There are several reasons for such differences: among others, the use of different distances or differences in the internal extinction. For example, James et al. (2004) adopted a constant value of 1.1 mag for the extinction of all the galaxies, regardless their morphological type or inclinations. This value is higher than the extinction we reported in our Section 3. Also, different number of H II regions detected might change the value of the SFR, as in UGC 11820, where in a previous work, up to 40 H II regions were considered for this galaxy, and a SFR of 2.9 $10^{-2}$ $M_\odot$ $yr^{-1}$ (Reyes-Pérez, 2009).

### 3.6. *Star Formation Efficiency*

Galaxies with distinct spiral arms, like Sc type, have larger SFR due to the fact that they have a large amount of gas and a prominent density wave (Kennicutt, 1998). Late-type galaxies (Sm, Im and dS) have, in general, smaller values of the SFR despite their large amount of gas. However, when the SFR surface density is considered, galaxies behave very differently. Hunter & Elmegreen (2004), using a sample of Sm, Im and Blue Compact Dwarf galaxies (BCD), found out that the latter were the galaxies with the largest SFR surface densities (see their figure 5). A similar result is found here, using the optical radius instead of the disk scale length in V which, according to Hunter & Elmegreen (2004), make no differences for SFR surface densities lower than 0.05 M$_\odot yr^{-1} kpc^{-2}$. The SFR surface density for dS galaxies is larger (with an average value of 1.6 10$^{-4}$ M$_\odot$ $yr^{-1} kpc^{-2}$) than that of Sm galaxies (on average 7.7 10$^{-5}$ M$_\odot$ $yr^{-1} kpc^{-2}$). So, dS are forming a larger number of stars per area than Sm, despite the fact that the latter on average has twice more gas mass than dS (12.5 $M_\odot$ $yr^{-1}$ in dS vs. 8.5 $M_\odot$ $yr^{-1}$ in Sm).

In order to study this effect, the well-known Kennicutt-Schmidt law is applied to our data and is presented in Figure 5 for both $H_\alpha$ (top) and $FUV$ (bottom). As no information is available on the amount of molecular gas in these galaxies, the gas density was determined as $1.3 \times \Sigma(HI)$, the ratio between the H I and the molecular gas being constant (Leroy et al. 2005; Saintonge et al. 2011). We plot the efficiency lines (defined as the ratio between the $SFR$ and the gas mass) at 1, 10 and 100%, as in Kennicutt (1998). From the results of Figure 5, it cannot be concluded that dS are more efficient forming stars than Sm or vice versa. Most of the galaxies in the sample are near the 10% efficiency line for both calibrators, although $FUV$ values are located mostly between the 10% and 100%, while there are galaxies between the 1% and 10% line with $H_\alpha$. The galaxies UGC 3086 and UGC 3778 seem to be very efficient, closest to the 100% efficiency line.

If the figure was divided into three different regions according to the efficiency, the number of Sm and dS galaxies would be very similar in all of them, but at intermediate/lower efficiency the Sm type are slightly more numerous.

The larger amount of gas in Sm with a similar $HI$ mass might indicate that they will form stars for a longer time than dS, or that they have started to form stars more recently than dS, but we do not have yet enough information to discriminate between these two options.

## 4. SPATIAL DISTRIBUTION OF THE STAR FORMATION

The spatial distribution of the H II regions inside late-type galaxies was studied in previous investigations (e.g. Hodge 1969; Hunter & Gallagher 1986; Brosch et al. 1998; Sánchez & Alfaro 2008). Roye & Hunter (2000) concluded that, in irregular galaxies, the distribution of H II regions is mostly random, although the majority of the regions are concentrated in the central part of the galaxy. This is very similar to the previous results of Hodge (1969) and Brosch et al. (1998) who got a global asymmetry in the H II regions locations, with a concentration towards the center of the galaxies. On the contrary, Hunter (1982) only found a random distribution of H II regions in late-type galaxies, with the exception of a few chains. Elmegreen & Elmegreen (1980) found out that the largest H II regions in barred galaxies are located at the end of the bar. A study of the distribution of the H II regions within galaxies in our sample might help shed some light on the previous discussion. Moreover, the locations of the H II regions inside a galaxy might give information about possible previous interactions between galaxies.

In order to study if there is any particular distribution in the H II regions locations for the galaxies of our sample, we selected those galaxies with more than 10 H II regions (17 galaxies, 10 Sm and 7 dS) and inclinations lower than $70^o$ to avoid problems of resolution and projection. From a quick inspection of the images ($H_\alpha$ shown in figure ?? and broadband images not shown here) it can be concluded that only DDO 36, NGC 2552, UGC 6151 and UGC 6713 might have spiral arms because the H II regions are very much aligned along this structure of the galaxy. The regions inside UGC 6399 and UGC 891 are also aligned along the major axis, but as these galaxies have large inclinations, it might be just a projection effect. For the other 11 galaxies, the H II regions are randomly distributed.

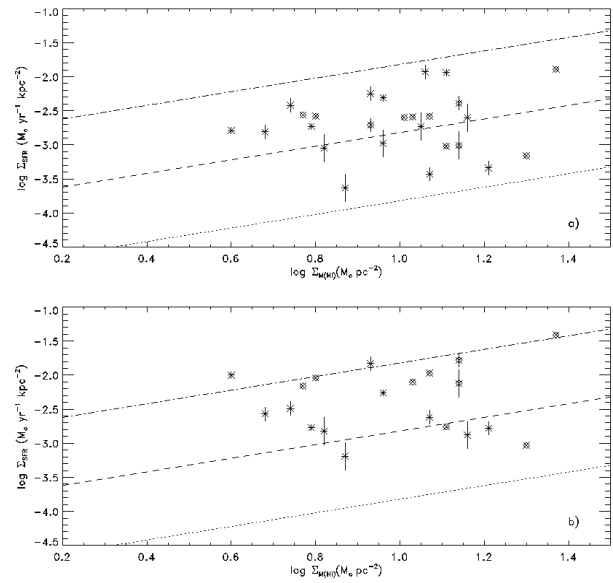

Fig. 5. Correlation between the SFR surface density and the average surface densities of $H_I$. The dashed lines correspond to constant global Star Formation Efficiencies (1, 10, 100%) from Kennicutt (1998). The Sm and dS galaxies are shown with different symbols, stars and diamonds respectively. The graph at the top shows the KS law when the star formation was determined from the $H_\alpha$ luminosity, while the one at the bottom shows the star formation determined with the FUV continuum.

### 4.1. *Concentration parameter*

One parameter that can give us important information about the distribution of the star formation within galaxies is the concentration index (hereafter $CI_n$), which is defined as the ratio between the number of H II regions in the inner part and outer part of the galaxy (Roye & Hunter 2000). With this definition of the $CI_n$, a value of 0 means that all H II regions are in the outer part of the galaxy, and the larger the $CI_n$ the more concentrated towards the center the regions are. In the determination of the $CI_n$ for the 17 galaxies selected before we follow Roye & Hunter (2000); the values are listed in column 2 of Table 4. All the galaxies have $CI_n$s larger than 1, and four of them have $CI_n$ values larger than 10, indicating a large concentration of regions towards the center. Only six of the galaxies have CI values smaller than 4, so their H II regions are located mostly outside the inner part. In general, it can be said that the galaxies in this subsample do not have their H II regions in their outskirts.

Another CI can be defined using the flux of the H II regions ($F_{in}/F_{out}$) instead of the number of them (hereafter, $CI_f$). This is presented in column 3 of Table 4. Again, large values of $CI_f$ indicate a large amount of flux towards the center, and small values indicate a large amount of flux from the outer regions. Eight of the galaxies have $CI_f$ smaller than 4, therefore their flux come mostly from the outskirts. On the contrary, five galaxies have a strong luminosity from their center.

The most interesting result comes when we compare both CIs. One might think that their values should be similar, e.g. if a galaxy have most of its H II regions in the inner part, then most of the flux of the galaxy (in $H_\alpha$) should come from it as well. Thereupon, it could be said that the galaxy is "well balanced": there is more flux because there are more H II regions. Such comparison can be done from columns 2 and 3 of Table 4. Half of the sample have similar values of both concentration index. However, there are three galaxies (UGC 4121, UGC 6151, and UGC 6840) which have values of their $CI_n$ much higher (a factor 1.5 or larger) than the $CI_f$. Therefore these galaxies have a large number of regions inside their central region, but they are not very luminous. On the contrary, four of the galaxies in the sample (24 %) have a $CI_f$ much larger than the $CI_n$, indicating that, despite being less in number, the regions in the center of the galaxy are brighter than those in the outskirts.

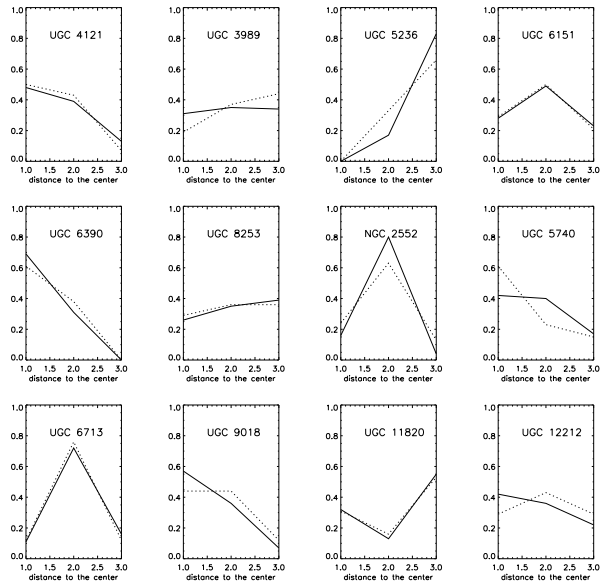

Fig. 6. Distribution of the number of H II regions (solid line) and their fluxes (dotted lines) along the distance to the center of the regions, for the non-barred galaxies. The y axis is the normalized number of H II regions (solid) or flux (dotted) of the galaxy and the x axis is the location of the galaxy in the any of the three regions we divided the galaxies into: inner, middle and outer part. Each galaxy has a different trend: some galaxies have more regions in the inner part, some others in the outer and some others in the middle. UGC 3989 has its star-forming regions quite homogeneously distributed along the radius, but the flux increases towards the edge. For most of the galaxies, the number of regions and the associated fluxes mimic each other.

It can be concluded that the galaxies in this subsample have a larger number of H II regions in the internal half of the galaxy, which are also more luminous than the regions in the external half of the galaxy. However, this is intriguing because in a visual inspection of the galaxies (see figure 6), the large number of the central H II regions and their dominance of the luminosity of the galaxy is not clear. Therefore, we divided the galaxy into three concentric regions: the inner region with a radius of 1/3 of the total optical radius, the middle region, a torus between 1/3 and 2/3 of the $r_{25}$, and the outer region, from there up to the optical radius. They are marked with circles in figure 6. The number of H II regions inside each of these regions, as well as their total flux are listed in Table 4, columns 4 to 15. There is not a clear pattern, as can be seen in figures 6 and 7, where the profiles of the number of H II regions (solid line) and the flux (dotted line), both normalized to the total number of regions (or flux) for each galaxy, are shown. The most peculiar profile might be the one of UGC 6840, with fewer regions in the inner annulus, dropping toward the middle section and increasing dramatically at the outer part of the galaxy. Moreover, this galaxy, along with UGC 3989, has the largest differences between the number and the flux distribution. Regarding the non barred galaxies, four out of 12 have the number and the flux of their H II regions concentrated in one of the annulus of the galaxy, while for other four the distribution is very uniform along the galactocentric distance. Four of the five barred galaxies have a large amount of H II regions at the middle part of the galaxy. These results do not support the idea that bars inhibit the star formation in the central part (e.g. James & Percival, 2018; Tubbs 1982; Kim et al. 2018) because only DDO 36 have fewer regions inside the central annulus. It is interesting to note that all galaxies with a clear spiral structure have all –or almost all of their star formation regions concentrated in the middle and outer sectors of the galaxy, following the spiral arms. On the other hand, the regions of star formation in galaxies without or with a not-so-well defined spiral structure do not show any special distribution.

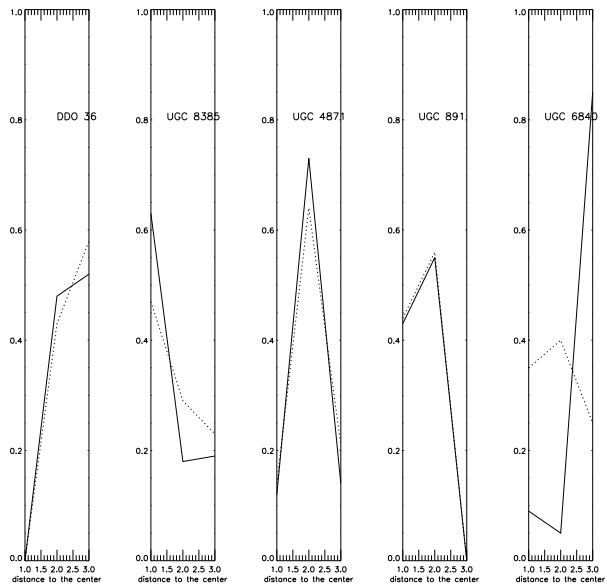

Fig. 7. Distribution of number of H II regions (solid line) and their fluxes (dotted line) along the radius for barred galaxies. Only two of them have most of their regions and flux at the middle part of the galaxy, where the bar is expected to end, and have important influence in the star formation. Axis as in figure 6. From left to right the galaxies are: DDO 36, UGC 8385, UGC 4871, UGC 891 and UGC 6840.

From these figures and the values in Table 4 it can be concluded that the H II regions of the galaxies in our sample show a variety of distributions, more similar to a random behavior than to a clear pattern, as observed in spiral galaxies.

TABLE 4

DISTRIBUTION OF THE STAR FORMATION WITH THE RADIUS

| Galaxy | $CI_n$ | $CI_f$ | Inner n | %n | $F_{H\alpha}\ 10^{-14}$ [erg s$^{-1}$ cm$^{-2}$] | %F | Middle n | %n | $F_{H\alpha}\ 10^{-14}$ [erg s$^{-1}$ cm$^{-2}$] | %F | Outer n | %n | $F_{H\alpha}\ 10^{-14}$ [erg s$^{-1}$ cm$^{-2}$] | %F |
|---|---|---|---|---|---|---|---|---|---|---|---|---|---|---|
| UGC4121 | 10.0 | 6.6 | 7 | 50.0 | 0.60 | 48.0 | 6 | 42.9 | 0.49 | 39.2 | 1 | 7.1 | 0.16 | 12.8 |
| UGC3989 | 2.0 | 3.8 | 3 | 18.8 | 0.37 | 31.1 | 6 | 37.5 | 0.42 | 35.3 | 7 | 43.7 | 0.40 | 33.6 |
| UGC5236 | 1.5 | 0.7 | 0 | 0.0 | 0.00 | 0.0 | 5 | 33.3 | 1.11 | 16.7 | 10 | 66.7 | 5.52 | 83.3 |
| UGC6151 | 7.2 | 1.7 | 4 | 28.6 | 0.34 | 28.0 | 7 | 50.0 | 0.59 | 48.9 | 3 | 21.4 | 0.28 | 23.1 |
| UGC6399 | 13.3 | 14.6 | 8 | 61.5 | 0.31 | 68.9 | 5 | 38.5 | 0.14 | 31.1 | 0 | 0.0 | 0.00 | 0.0 |
| UGC8253 | 3.0 | 2.9 | 4 | 28.6 | 0.41 | 26.1 | 5 | 35.7 | 0.55 | 35.0 | 5 | 35.7 | 0.61 | 38.9 |
| DDO36 | 4.0 | 3.0 | 0 | 0.0 | 0.00 | 0.0 | 11 | 42.3 | 1.05 | 47.9 | 15 | 57.7 | 1.14 | 52.1 |
| NGC2552 | 8.8 | 21.4 | 15 | 23.8 | 15.97 | 15.7 | 40 | 63.5 | 81.43 | 80.2 | 8 | 12.7 | 4.18 | 4.1 |
| UGC8385 | 5.1 | 9.0 | 8 | 47.1 | 5.25 | 62.9 | 5 | 29.4 | 1.53 | 18.3 | 4 | 23.5 | 1.57 | 18.8 |
| UGC4871 | 10.0 | 13.0 | 2 | 14.3 | 0.17 | 12.2 | 9 | 64.3 | 1.02 | 73.4 | 3 | 21.4 | 0.20 | 14.4 |
| UGC5740 | 9.0 | 7.7 | 8 | 61.5 | 1.72 | 42.5 | 3 | 23.1 | 1.63 | 40.2 | 2 | 15.4 | 0.70 | 17.3 |
| UGC6713 | 3.6 | 3.0 | 2 | 11.8 | 0.49 | 11.0 | 13 | 76.4 | 3.24 | 72.5 | 2 | 11.8 | 0.74 | 16.5 |
| UGC9018 | 8.8 | 19.2 | 7 | 43.8 | 3.24 | 56.6 | 7 | 43.8 | 2.06 | 36.1 | 2 | 12.5 | 0.42 | 7.3 |
| UGC11820 | 2.9 | 2.9 | 6 | 31.6 | 1.71 | 32.3 | 3 | 15.8 | 0.67 | 12.6 | 10 | 52.6 | 2.92 | 55.1 |
| UGC12212 | 3.0 | 4.5 | 6 | 28.6 | 8.47 | 41.9 | 9 | 42.8 | 7.21 | 35.6 | 6 | 28.6 | 4.55 | 22.5 |
| UGC891 | 17.3 | 25.2 | 7 | 43.8 | 3.58 | 43.3 | 9 | 56.2 | 4.69 | 56.7 | 0 | 0.0 | 0.00 | 0.0 |
| UGC6840 | 6.0 | 0.63 | 7 | 35.0 | 0.26 | 9.2 | 8 | 40.0 | 0.15 | 5.3 | 5 | 25.0 | 2.41 | 85.5 |

Galaxy name is in column 1. In columns 2 and 3, the concentration rates are listed by number ($CI_n$) and by flux ($CI_f$), respectively. In columns 4 and 5 are the number of H II regions and their percentages. In columns 6 and 7 are the total $H_\alpha$ flux of these regions and their percentage, within the innermost section of the galaxy. This section corresponds to a disk with a radius equal to one-third of $R_{25}$. In columns 8 and 9 are the number of H II regions and the percentage to these. In columns 10 and 11 are the total $H\alpha$ flux within the middle section of the galaxy and the corresponding percentage. This section corresponds to a ring with an outer radius equal to two-thirds to $R_{25}$. In columns 12 and 13 are the number of H II regions and the percentage corresponding to these. In columns 14 and 15 are the total $H\alpha$ flux within the outer section of the galaxy and their percentage. This section corresponds to a torus with an outer radius equal to $R_{25}$.

TABLE 5

DISTRIBUTION OF STAR FORMATION IN ORIENTATION

| Galaxia | NE n | %n | $F_{H\alpha}\ 10^{-14}$ [erg s$^{-1}$ cm$^{-2}$] | %F | NW n | %n | $F_{H\alpha}\ 10^{-14}$ [erg s$^{-1}$ cm$^{-2}$] | %F | SE n | %n | $F_{H\alpha}\ 10^{-14}$ [erg s$^{-1}$ cm$^{-2}$] | %F | SW n | %n | $F_{H\alpha}\ 10^{-14}$ [erg s$^{-1}$ cm$^{-2}$] | %F | $AI_n$ | $AI_f$ |
|---|---|---|---|---|---|---|---|---|---|---|---|---|---|---|---|---|---|---|
| UGC4121 | 3 | 21.4 | 0.26 | 21.8 | 5 | 35.8 | 0.46 | 38.7 | 3 | 21.4 | 0.27 | 22.7 | 3 | 21.4 | 0.20 | 16.8 | 0.75 | 0.65 |
| UGC3989 | 4 | 25.0 | 0.22 | 18.6 | 4 | 25.0 | 0.37 | 31.4 | 4 | 25.0 | 0.37 | 31.4 | 4 | 25.0 | 0.22 | 18.6 | 1.00 | 0.75 |
| UGC5236 | 1 | 6.7 | 0.24 | 3.6 | 6 | 40.0 | 2.65 | 40.0 | 2 | 13.3 | 2.80 | 42.2 | 6 | 40.0 | 0.94 | 14.2 | 0.50 | 0.85 |
| UGC6151 | 2 | 14.3 | 0.16 | 13.5 | 2 | 14.3 | 0.18 | 15.0 | 5 | 35.7 | 0.45 | 37.4 | 5 | 35.7 | 0.41 | 34.1 | 0.56 | 0.56 |
| UGC6399 | 2 | 15.3 | 0.12 | 26.0 | 5 | 38.5 | 0.17 | 37.0 | 6 | 46.2 | 0.17 | 37.0 | 0 | 0.0 | 0.00 | 0.0 | 0.86 | 0.84 |
| UGC8253 | 5 | 35.7 | 0.53 | 34.0 | 2 | 14.3 | 0.18 | 11.5 | 4 | 28.6 | 0.49 | 31.4 | 3 | 21.4 | 0.36 | 23.1 | 0.75 | 0.65 |
| DDO36 | 7 | 26.9 | 0.44 | 20.1 | 2 | 7.7 | 0.34 | 15.5 | 9 | 34.6 | 0.46 | 21.0 | 8 | 30.8 | 0.95 | 43.4 | 1.00 | 0.90 |
| NGC2552 | 12 | 19.0 | 22.40 | 22.2 | 11 | 17.5 | 16.46 | 16.4 | 17 | 27.0 | 31.15 | 31.0 | 23 | 36.5 | 30.58 | 30.4 | 0.98 | 0.98 |
| UGC8385 | 3 | 17.6 | 0.76 | 9.1 | 5 | 29.5 | 1.89 | 22.6 | 4 | 23.5 | 1.18 | 14.1 | 5 | 29.4 | 4.52 | 54.2 | 0.89 | 0.46 |
| UGC4871 | 4 | 28.6 | 0.26 | 18.7 | 5 | 35.7 | 0.62 | 44.6 | 2 | 14.3 | 0.17 | 12.2 | 3 | 21.4 | 0.34 | 24.5 | 0.56 | 0.57 |
| UGC5740 | 1 | 7.7 | 0.23 | 5.7 | 7 | 53.8 | 2.75 | 67.9 | 3 | 23.1 | 0.53 | 13.1 | 2 | 15.4 | 0.54 | 13.3 | 0.86 | 0.87 |
| UGC6713 | 4 | 23.6 | 0.41 | 9.2 | 7 | 41.2 | 2.59 | 57.9 | 3 | 17.6 | 0.41 | 9.2 | 3 | 17.6 | 1.06 | 23.7 | 0.70 | 0.55 |
| UGC9018 | 5 | 31.3 | 1.20 | 20.9 | 2 | 12.4 | 0.45 | 7.9 | 4 | 25.0 | 1.73 | 30.2 | 5 | 31.3 | 2.35 | 41.0 | 0.78 | 0.57 |
| UGC11820 | 5 | 26.3 | 1.58 | 30.2 | 10 | 52.6 | 2.72 | 52.0 | 1 | 5.3 | 0.21 | 4.0 | 3 | 15.8 | 0.72 | 13.8 | 0.90 | 0.73 |
| UGC12212 | 5 | 23.8 | 4.56 | 22.3 | 2 | 9.5 | 3.36 | 16.5 | 6 | 28.6 | 6.01 | 29.5 | 8 | 38.1 | 6.46 | 31.7 | 0.91 | 0.88 |
| UGC891 | 9 | 56.2 | 5.14 | 62.1 | 1 | 6.3 | 0.00 | 0.0 | 1 | 6.3 | 1.22 | 14.8 | 5 | 31.2 | 1.91 | 23.1 | 0.78 | 0.52 |
| UGC6840 | 7 | 35.0 | 0.24 | 8.6 | 3 | 15.0 | 0.02 | 0.7 | 3 | 15.0 | 0.07 | 2.5 | 7 | 35.0 | 2.46 | 88.2 | 1.00 | 0.12 |

Galaxy name is in column 1. In columns 2 and 3 are the number of H II regions within the northeastern quadrant (NE) and percentage corresponding to these. In column 4 is the total $H_\alpha$ flux of these regions, and its corresponding percentage is in column 5. In the same way, the number of H II regions (n), the corresponding percentage (%n), the total $H\alpha$ flux of these regions and its percentage (%F), are presented for the quadrants northwest (NW), southeast (SE) and southwest (SW). The asymmetry indices by number ($AI_n$) and by flux ($AI_f$) are shown in columns 18 and 19.

4.2. *Asymmetry in the H II regions distribution*

In a recent study of the dwarf, interacting spiral galaxy IC 1727, it was found out that the largest and brightest of its H II regions are facing the companion galaxy, NGC 672 (Ramirez-Ballinas & Hidalgo-Gámez 2014). We aim to check if the H II were located in a particular zone in the galaxy, because such accumulation in the star formation onto one particular direction might indicate an interaction. There are two possible approaches, the asymmetry index AI, (Roye & Hunter, 2000) or a more simple one: to just check if there is any preferred locus of star formation inside each galaxy, dividing these into the four orientations as it was done in IC 1727. We will first consider the latter approach. In order to do it, the 17 galaxies studied at the previous subsection were divided into four directions: NE, NW, SE and SW (the North is up and East to the left). The values of the number of H II regions and the fluxes for each quadrant are listed in Table 5. Half of the galaxies in the sample have a quadrant (or two) with a meaningful increase in the number of H II regions (larger than 35%). This might indicate a previous interaction. Moreover, two of them, both barred galaxies, have the majority of regions at opposite quadrants, while another four galaxies have them in adjacent quadrants. A likely explanation might be that such star formation is due to gas coming out along the bar, as it is seen in other galaxies (Sérsic & Pastoriza 1967 ; Ho et al. 1997).

The asymmetry index, based on the number of H II regions ($AI_n$), is defined as the ratio between the number of regions in the poorest side and the richest side. An AI by number (or by flux, see below) equal to one means that the galaxy is perfectly symmetric along that axis (major or minor axis), that is, we have the same number of H II regions on one side as the other. On the contrary, an AI of 0 means that all H II regions are concentrated on only one side of the galaxy. We used our images in V and R to determine the major and minor axes and count the number of H II regions at each side of the axes. We decided to proceed as in Roye & Hunter (2000), where they a posteriori chose the axis (major or minor) with the largest symmetry. The $AI_n$ values are shown in column 18 of Table 5. We found that 80% of the galaxies in the sample have an $AI_n \geq 0.75$ and three of them are perfectly symmetric ($AI_n = 1$).

As previously done with the CI, an asymmetry index based on the flux can be determined as the ratio between the flux of all the regions in one side and the flux in the other side ($AI_f$). This index is shown in column 19 of Table 5. Again, our galaxies appear to be very symmetrical because only two of them have $AI_f < 0.5$. According to the AI, most of the galaxies in our sample are predominantly symmetric in both the distribution of the H II regions, and the luminosity of them.

## 5. IS A STARBURST ALWAYS A STARBURST?

Several authors (e.g. Kennicutt 1998; Lee et al. 2009; Gilbank et al. 2010) claimed that the $H_\alpha$ photons from outside the H II regions, the so-called Diffuse Ionized Gass (hereafter, DIG), can also be used in the SFR determination because this gas is ionized by photons created by OB stars and they are coming out from the H II regions, as the nebulae are density bounded instead of radiation bounded. However, this is not always the case. There are many other processes which can ionize this gas which are not related to young stellar population, as shock waves (Rand 1998), turbulent mixing layers (Slavin et al. 1993), hot low-mass evolved stars (Flores-Fajardo et al. 2011), and radiation from WR stars (Hidalgo-Gámez 2005).

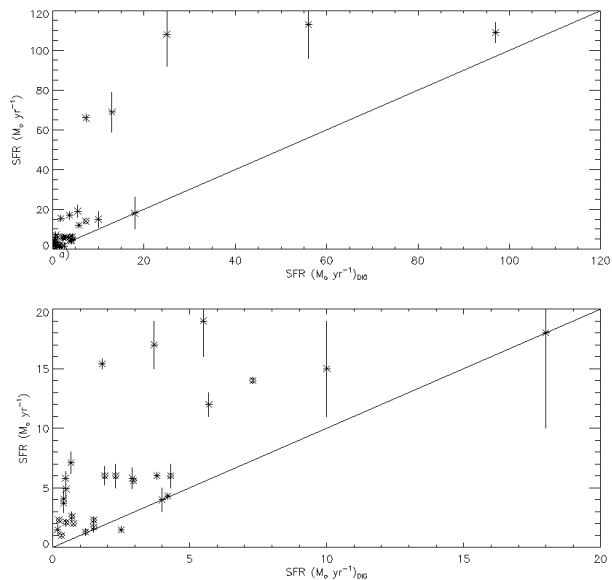

Fig. 8. Comparison of the SFR values when all the H$_\alpha$ photons are considered and when a limiting flux in H$_\alpha$ is considered. Bottom panel is a zoom to the lowest flux galaxies.

It is not easy to know which is the most important ionizing source for a particular galaxy and how much of the H$_\alpha$ luminosity comes from each of the sources using only H$_\alpha$ images. If other processes are at work at the ionization of the DIG, the SFR might be overestimated for those galaxies with a large amount of DIG. One example might be the irregular dwarf galaxy IC 10, which is considered to be a starburst based on its colour (Richer et al. 2001), the high content of WR stars (Massey & Holmes 2002) and the large number of H II regions (Hodge & Lee 1990), with a SFR of about $1 M_\odot \ yr^{-1}$ (Zucker 2005). However, there is a strong emission from the DIG inside this galaxy, and up to 50% of this DIG is ionized by sources other than leaked photons (Hidalgo-Gámez 2005). Therefore, IC 10 might not be considered as a starburst galaxy (Hidalgo-Gámez & Magaña-Serrano 2017). The inclusion of the DIG within the H$_\alpha$ luminosity used in the SFR should overestimate the values and it can be risky unless it is well known that leaked photons are the only responsible source of the ionization of this gas.

Unfortunately, the amount of DIG cannot be known in advance for a particular galaxy. Then, we decided to use previous studies on the DIG in late-type galaxies (e.g. Hidalgo-Gámez 2006; Hidalgo-Gámez, 2007; Hidalgo-Gámez 2005; Hidalgo-Gámez & Peimbert 2007) to determine the boundary flux between an H II region and the DIG. We estimate this flux to be of the order of $10^{-17} erg\ cm^{-2} s^{-1}$. Then, we extracted the part of the $H_\alpha$ emission with larger fluxes than this value and a new SFR was determined from the new luminosities. The results are listed in column 10 of Table 3. A quick comparison can be done between columns 6 and 10 and it is shown in figure 8. The number of galaxies with differences in the SFR when DIG is not included is quite large. For a total of 21 out of 36 of the galaxies in the sample, the SFR without the DIG is 50% lower than with it. There is no real differences between Sm (64%) and dS (53%). For seven of these galaxies the differences in the SFR with and without DIG is of 90%. Probably, as these are late-type galaxies with a large amount of gas, the contribution of the gas between the H II regions to the Star Formation Rate is larger and, before including it in the calculation, a careful study on the ionization source of such gas should be done.

## 6. CONCLUSIONS

In this work, the Star Formation Rates for a sample of late-type galaxies, Sm and dwarf spirals, have been determined. It is the first estimation of the star formation rate for more than half of the sample considered here. The SFR have been determined using the $H_\alpha$ flux for 36 galaxies and the FUV flux only for 30 of them. The fluxes used, both $H_\alpha$ and FUV, have been corrected for internal and external extinctions.

The bulk of the galaxies in our sample have SFRs lower than $0.1 M_\odot\ yr^{-1}$ for both diagnostic methods. These values are common for late-type galaxies. It is interesting to notice that SFR(FUV) is normally larger than the SFR determined with $H_\alpha$. However, the largest SFRs were obtained from $H_\alpha$ luminosities.

We also noticed a difference in the SFR between Sm and dS galaxies in the sense that the former had higher SFR values than dS with both methods.

If dS galaxies are not only smaller than Sm but a latter type of galaxies in between Sm and Irr galaxies, their lower SFR could be explained by an older star formation event for the dS galaxies. However, when the density of SFR ($SFR/kpc^2$) is considered, both type of galaxies have similar values. These are similar values to those found before (e.g. Hunter & Elmergreen 2004), although the dispersion in our sample is larger.

We studied the role played by the bars as drivers of the Star Formation in late-type galaxies and noticed that non-barred galaxies always have larger SFR than barred ones. Also, we studied the influence of the spiral arms in the SFR and found that, for late-type galaxies, the spiral wave might be not that important and other mechanisms might trigger the Star Formation, as in irregular galaxies.

Considering the Star Formation Efficiency, both type of galaxies have a similar efficiency of star formation despite of the greater amount of gas in the Sm and the differences in their gas mass densities.

We also studied the distribution of the H II regions for those galaxies in our sample with the largest number of regions. We used two approaches: the concentration and asymmetry indexes as in Roye & Hunter (2000) and a more detailed distribution with smaller divisions. We concluded that late-type galaxies are very symmetrical and have their regions quite concentrated. Moreover, although there is a gradient in the number and the fluxes of H II regions for more than half of the galaxies studied, such gradient is not unique, being positive or negative for some galaxies and even with two slopes in others.

Finally, we noticed that the inclusion of the DIG within the estimation of $H_\alpha$ luminosity might increase SFR values more than previously thought. Therefore, the $H_\alpha$ luminosity should not be included unless it is well known that leaking photons from the H II regions are the only source of the ionization.

## 7. ACKNOWLEDGEMENTS

The authors wish to thank an anonymous referee for comments which have improved the manuscript. This investigation has been supported by the Instituto Politecnico Nacional (México) under the research projects SIP-20196007, SIP-20181188 and SIP-20196007. The authors wish to thank Mr. P. Garcés for his careful revision of the manuscript. Part of this work was done while one of the authors (M.A.M.S) was at ESO/Chile for a short-term visit. This work is part of M.A. Magaña-Serrano's Ph.D. thesis. This research has made use of the NASA/IPAC Extragalactic Database (NED) which is operated by the Jet Propulsion Laboratory, California Institute of Technology, under contract with the National Aeronautics and Space Administration. Based upon observations carried out at the Observatorio Astronómico Nacional on the Sierra San Pedro Mártir (OAN-SPM), Baja California, México.

## REFERENCES

Almoznino, E., & Brosch, N. 1996, MNRAS, 31, 29
Audcent-Ross, F. M., Meurer, G. R., Wong, O. I., et al. 2018, MNRAS, 480, 119
Bell, E. F., & Kennicutt, R. C. 2001, Galaxy Disks and Disk Galaxies, 230, 305
Boselli, A., Boissier, S., Cortese, L., et al. 2009, ApJ, 706, 1527
Boselli, A., Fossati, M., Gavazzi, G., et al. 2015, A&A, 579, A102
Brosch, N., Heller, A., & Almoznino, E. 1998, MNRAS, 300, 1091
Buat, V., Takeuchi, T. T., Burgarella, D., Giovannoli, E., & Murata, K. L. 2009, A&A, 507, 693
Calzetti, D. 1997, , 408, 403
Calzetti, D. et al. 2000, ApJ, 533, 682
Calzetti, D. 2013, Secular Evolution of Galaxies: Cambridge University Press, 419
Cardelli, J. A., Clayton, G. C., & Mathis, J. S. 1989, ApJ, 345, 245
Clayton, G. C., & Cardelli, J. A. 1988, AJ, 96, 695
Danieli, S., van Dokkum, P., Merritt, A., et al. 2017, ApJ, 837, 136

Dekel, A., Silk, J. 1986, ApJ, 303, 39

D'Ercole, A., Brighenti, F. 1999, MNRAS, 309, 941

Elmegreen, D. M., & Elmegreen, B. G. 1980, AJ, 85, 1325

Flores-Fajardo, N., Morisset, C., Stasińska, G., & Binette, L. 2011, MNRAS, 415, 2182

Gilbank, D. G., Baldry, I. K., Balogh, M. L. et al. 2010, MNRAS, 405, 2594

Hidalgo-Gámez, A. M. 2004, RevMex A&A, 40, 37

Hidalgo-Gámez, A. M. 2005, A&A, 442, 443

Hidalgo-Gámez, A. M. 2007, AJ, 134, 1447

Hidalgo-Gámez, A. M. 2006, AJ, 131, 2078

Hidalgo-Gamez, A. M.& Magana-Serrano, M.A. E.2017, Multi-Scale Star Formation, IRyA-UNAM

Hidalgo-Gámez, A. M., Moranchel-Basurto, A., & González-Fajardo, A. F. 2012, RevMex A&A, 48, 183

Hidalgo-Gámez, A. M., & Peimbert, A. 2007, AJ, 133, 1874

Ho, L., Filippenko, A., & Sargent, W. 1997 ApJ, 4897, 591

Hodge, P. W. 1969, ApJ, 156, 847

Hodge, P., & Lee, M. G. 1990, PASP, 102, 26

Huchtmeier, W.K. & Richter, O.-G 1989, A&A, 201, 1

Hunter, D. A. 1982, ApJ, 260, 81

Hunter, D. A., & Elmegreen, B. G. 2004, AJ, 128, 2170

Hunter, D. A. & Gallagher, J. S., III 1986, PASP, 98, 5

Hunter, D. A., Elmegreen, B. G., & Ludka, B. C. 2010, AJ, 139, 447

Iglesias-Páramo, J., Boselli, A., Gavazzi, G., & Zaccardo, A. 2004, A&A, 421, 887

James, P. A., & Percival, S. M. 2018, MNRAS, 474, 3101

James, P. A., Prescott, M., & Baldry, I. K. 2008, A&A, 484, 703

James, P. A., Shane, N. S., Beckman, J. E., et al. 2004, A&A, 414, 23

Jog, C. J. 2013, MNRAS, 434, 56

Karachentsev, I. D., & Kaisina, E. I. 2013, AJ, 146, 46

Kennicutt, R. C., Jr. 1998, ApJ, 498, 541

Kennicutt, R. C., Jr. 1998, ARA&A, 36, 189

Kim, E., Hwang, H. S., Chung, H., et al. 2017, ApJ, 845, 93

Kim, E., Kim, S. S., Choi, Y.-Y. et al. 2018, MNRAS, 479, 562

Lee, J. C., Gil de Paz, A., Tremonti, C., et al. 2009, ApJ, 706, 599

Lee, J. C., Gil de Paz, A., Tremonti, C., Kennicutt, R., van Zee, L., & LVL Team. 2007, American Astronomical Society, AAS Meeting 211, Bulletin of AAS, 39, 895

Leroy, A., Bolatto, A. D., Simon, J. D., & Blitz, L. 2005, ApJ, 625, 763

López F. R. et al. 2018, A&A, 615, 27

Martin, D. C. & GALEX Science Team. 2005, BAAS, 207, 5201

Melioli, C., Brighenti, F., D'Ercole, A. 2015, MNRAS, 446, 299

Marzke, R. O., & da Costa, L. N. 1997, AJ, 113, 185

Massey, P., & Holmes, S. 2002, ApJ, 580, L35

Mateo, M. L. 1998, ARA&A, 36, 435

Morrissey, P. & GALEX Science Team. 2005, BAAS, 37 , 1454

Morrissey, P. & GALEX Science Team. 2005, ApJS,173 , 682

Ramirez-Ballinas, I., & Hidalgo-Gámez, A. M. 2014, MNRAS, 442, 2282

Rand, R. J. 1998, ApJ, 501, 137

Reyes-Pérez. 2009. "Tasa de Formación Estelar y Funciónes de Luminosidad de galaxias enanas" tesis de licenciatura, Escuela Superior de Física y Matemáticas-IPN.

Richer, M. G., Bullejos, A., Borissova, J., et al. 2001, A&A, 370, 34

Riess, A. G., Macri, L., Casertano, S., et al. 2011, ApJ, 730, 119

Robitaille, T. P., & Whitney, B. A. 2010, ApJ, 710, L11

Rosa-González, D., Terlevich, E., & Terlevich, R. 2002, MNRAS, 332, 283

Rosenberg, J. L., Wu, Y., Le Floc'h, E., et al. 2008, ApJ, 674, 814

Roye, E. W., & Hunter, D. A. 2000, AJ, 119, 1145

Saintonge, A., Kauffmann, G., Wang, J., et al. 2011, MNRAS, 415, 61

Salim, S., Rich, R. M., Charlot, S., et al. 2007, ApJS, 173, 267

Sánchez, N.,& Alfaro, E. J. 2008, ApJS, 178, 1

Segura et al. 1996, A&A, 316, 133

Sérsic, J. L., & Pastoriza, M. 1967, PASP, 79, 152

Slavin, J. D., Shull, J. M., & Begelman, M. C. 1993, ApJ, 407, 83

Sullivan, M., Treyer, M. A., Ellis, R. S., & Mobasher, B. 2004, MNRAS, 350, 21

Ryder, S. D., & Dopita, M. A. 1994, ApJ, 430, 142

Tinsley B. M. 1981, ApJ, 250, 758

Tsai, C., Turner, J. L., Beck, S. C. et al. 2013, ApJ, 776, 70

Tubbs, A. D. 1982, ApJ, 255, 458

Van Zee, L. AJ, 121, 2003

TWang, Weichen et al. 2018, ApJ, 869, 161

Werk, J. K., Putman, M. E., & Meurer, G. R., PASP, 423, 287

Zucker, D. 2005, Starbursts: From 30 Doradus to Lyman Break Galaxies, 329, P87

Departamento de Física, ESFM-IPN, Mexico city, Mexico (mmaganas1500@alumno.ipn.mx, ahidalgo@esfm.ipn.mx).